\documentclass[cits,pdftex]{PoS}

\usepackage[utf8]{inputenc}
\usepackage{amsmath}
\usepackage{booktabs}
\usepackage{url}

\newcommand{\ee}{{\mathrm e}}
\newcommand{\Tr}{{\mathrm{Tr}}}
\renewcommand{\Re}{{\mathrm{Re}}}
\newcommand{\ket}[1]{{\left|#1\right\rangle}}
\newcommand{\bracket}[3]{{\left\langle#1\left|#2\right|#3\right\rangle}}
\newcommand{\expectationvalue}[1]{{\left\langle#1\right\rangle}}
\newcommand{\order}[1]{{\mathrm O(#1)}}
\newcommand{\PCAC}{{\mathrm{PCAC}}}
\newcommand{\ca}{{c_\mathrm A}}
\newcommand{\za}{{Z_\mathrm A}}
\newcommand{\csw}{{c_\mathrm{SW}}}
\newcommand{\fa}{{f_\mathrm A}}
\newcommand{\fp}{{f_\mathrm P}}

\title{Determination of $\ca$ in three-flavour lattice QCD with Wilson fermions and tree-level improved gauge action}

\ShortTitle{$\ca$ with $N_\mathrm f=3$ Wilson fermions and tree-level improved gauge action}

\author{%
  \includegraphics[width=2.5cm]{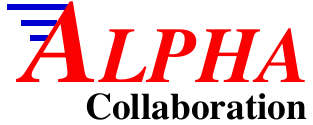}%
  \hfill%
  \raise22.35122pt\hbox{\footnotesize\it MS-TP-13-31}%
  \bigskip}

\author{John Bulava\\
  Trinity College Dublin\\
  College Green, Dublin 2, Ireland\\
  E-mail: \email{john.bulava@gmail.com}}

\author{Michele Della Morte\\
  IFIC (CSIC) Valencia\\
  c/ Catedrático José Beltrán, 2, 46980 Paterna, Spain\\
  E-mail: \email{dellamor@ific.uv.es}}

\author{Jochen Heitger\\
  Westfälische Wilhelms-Universität Münster, Institut für Theoretische Physik\\
  Wilhelm-Klemm-Straße 9, 48149 Münster, Germany\\
  E-mail: \email{heitger@uni-muenster.de}}

\author{\speaker{Christian Wittemeier}\\
  Westfälische Wilhelms-Universität Münster, Institut für Theoretische Physik\\
  Wilhelm-Klemm-Straße 9, 48149 Münster, Germany\\
  E-mail: \email{christian.wittemeier@uni-muenster.de}}

\abstract{We report on an ongoing non-perturbative determination of the improvement coefficient of the axial current, $\ca$, with three flavours of dynamical $\order{a}$ improved Wilson quarks and tree-level Symanzik improved gauge action.  Our computations are based on simulations with the openQCD code. The improvement condition for a range of couplings is formulated with Schrödinger functional boundary conditions and imposed along a line of constant physics in parameter space.  Our analysis involves correlation functions with boundary wave functions such that a large sensitivity to $\ca$ can be reached by exploiting the PCAC relation with two different pseudoscalar states.}

\FullConference{31st International Symposium on Lattice Field Theory LATTICE 2013\\
  July 29 -- August 3, 2013\\
  Mainz, Germany}

\begin{document}

\section{Introduction}

The Wilson discretization of the fermionic part of the QCD action introduces errors linear in the lattice spacing~$a$.  In Symanzik's improvement programme \cite{impr:sym1} these are reduced to $\order{a^2}$ by adding a dimension-five term to the action, the Sheikholeslami--Wohlert term, and dimension-four terms to the quark bilinears.  In particular, the axial current
\begin{align}
  A_\mu^a(x) = \bar\psi(x)T^a\gamma_\mu\gamma_5\psi(x)
\end{align}
(with $T^a$ acting in flavour space) is improved in the case of massless quarks by a term proportional to the derivative of the pseudoscalar density~$P$:
\begin{gather}
  (A_\mathrm I)_\mu^a(x) = \bar\psi(x)T^a\gamma_\mu\gamma_5\psi(x)+\ca\cdot{\textstyle\frac{1}{2}}\left(\partial_\mu+\partial^*_\mu\right)P(x), \qquad
  P(x) = \bar\psi(x)\gamma_5\psi(x), \\
  \partial_\mu f(x) = {\textstyle\frac{1}{a}}\left[f(x+a\hat\mu)-f(x)\right], \qquad
  \partial^*_\mu f(x) = {\textstyle\frac{1}{a}}\left[f(x)-f(x-a\hat\mu)\right].
\end{gather}

The axial current has various applications, such as the computation of PCAC quark masses or pseudoscalar decay constants.  It is particularly important for setting the scale via $f_\mathrm K$ as done in \cite{Fritzsch:2012wq}.  As a step to extend this programme to the three-flavour theory, we here present our calculation of $\ca$ with tree-level improved gauge action.  The improvement coefficient $\ca$ of the axial current has already been determined non-perturbatively in the quenched case (see e.g.\ \cite{Luscher:1996ug}), for $N_\mathrm f=2$ in Wilson QCD and for $N_\mathrm f=3$ with Iwasaki gauge action \cite{DellaMorte:2005se,Kaneko:2007wh}.  Since improvement coefficients are affected by $\order{a}$ ambiguities, it turned out that for a given discretization their values can differ quite significantly depending on the improvement conditions chosen (see e.g.\ \cite{Luscher:1996ug} compared to \cite{Bhattacharya:2000pn,Collins:2001mm}).  However, to ensure that the ambiguities vanish smoothly in the continuum limit, we impose our improvement condition along a line of constant physics, LCP (see also \cite{impr:babp,HQET:pap2,impr:babp_nf2}).  This amounts to vary the lattice spacing while keeping all physical length scales fixed.

All improvement conditions that are in practical use are based on the PCAC mass:
\begin{align}
  m_\PCAC & = \frac{\bracket{\alpha}{\partial_\mu A_\mu^a(x)}{\beta}}{2\bracket{\alpha}{P^a(x)}{\beta}}.
\end{align}
In the continuum it is independent of the point~$x$ and the states $\ket{\alpha}$ and $\ket{\beta}$, since it is derived from an operator identity, but it is violated at $\order{a}$ on the lattice.  For improvement we can choose two sets of states (or points~$x$) and adjust $\ca$ such that the associated $\order{a}$ improved masses agree.

In \cite{Luscher:1996ug} $\ca$ was computed in the quenched theory.  There the states used in the improvement condition differed in their (Schrödinger functional) periodicity angles~$\theta$.  However, this choice of states is not suitable for $N_\mathrm f>0$, because it would require to generate a separate ensemble for each value of $\theta$.  Instead, we rely on the strategy proposed in the $N_\mathrm f=2$ case \cite{DellaMorte:2005se}, where the states in the improvement condition are characterized by different wave functions, which are meant to approximate the ground and first excited state.  This will be detailed in the following section.

The gauge action which we consider in this article is the tree-level improved or Lüscher--Weisz action \cite{Luscher:1985zq}.  In addition to the Wilson plaquette term it includes sums over loops involving next-to-nearest neighbours.  Here, we take into account plaquettes and double-plaquettes, i.e. straight $2\times1$ loops (type~1 in \cite{Luscher:1985zq}):
\begin{align}
  S[U] & = \frac{2}{g_0^2}\cdot\sum_{i=0}^1c_i\sum_{\mathcal C\in\mathcal S_i}\Re\Tr\left(1-U(\mathcal C)\right).
\end{align}
$\mathcal S_0$ and $\mathcal S_1$ are the sets of all plaquettes and double-plaquettes, with loops that differ in orientation only considered identical, $U(\mathcal C)$ is the product of the link variables along the loop $\mathcal C$, and the coefficients are $c_0=5/3$ and $c_1=-1/12$.  For this action, the improvement coefficient~$\csw$ has been non-perturbatively determined in \cite{Bulava:2013cta}.

\section{Improvement Condition}

Following \cite{DellaMorte:2005se}, our improvement condition is based on the PCAC quark mass, which on the lattice we can write as
\begin{gather}
  m(x;\alpha,\beta) = r(x;\alpha,\beta)+a\ca\cdot s(x;\alpha,\beta), \\
  \label{e:r_s_definition}
  r(x;\alpha,\beta) = \frac{\bracket{\alpha}{\frac{1}{2}(\partial_\mu+\partial^*_\mu)(A(x))^a_0}{\beta}}{2\bracket{\alpha}{P(x)^a}{\beta}}, \qquad
  s(x;\alpha,\beta) = \frac{\bracket{\alpha}{\partial_\mu\partial^*_\mu(P(x))^a}{\beta}}{2\bracket{\alpha}{P(x)^a}{\beta}}.
\end{gather}
Here, $\ket{\alpha}$ and $\ket{\beta}$ denote two arbitrary states.  To define $\ca$ we choose two such pairs of states, $\ket{\alpha}$, $\ket{\beta}$ and $\ket{\gamma}$, $\ket{\delta}$ and require the PCAC masses computed from them to be equal.  From this condition we can extract the improvement coefficient:
\begin{align}
  \label{e:ca_definition_general}
  \ca & = -\frac{1}{a}\cdot\frac{r(x;\alpha,\beta)-r(x;\gamma,\delta)}{s(x;\alpha,\beta)-s(x;\gamma,\delta)}.
\end{align}

We work in a Schrödinger functional (SF) setup and prepare the initial states $\ket{\beta}$ and $\ket{\gamma}$ by applying a pseudoscalar two-quark operator $O^a(\omega)$ with suitable wave functions (see below).  Hence, we consider the correlators
\begin{gather}
  \fa(x_0;\omega) = -\frac{a^3}{3L^6}\sum_{\vec x}\expectationvalue{A^a_0(x)O^a(\omega)}, \qquad
  \fp(x_0;\omega) = -\frac{a^3}{3L^6}\sum_{\vec x}\expectationvalue{P^a(x)O^a(\omega)}, \\
  O^a(\omega) = a^6\sum_{\vec x\vec y}\bar\zeta(\vec x)\cdot T^a\gamma_5\cdot\omega(\vec x-\vec y)\cdot\zeta(\vec y),
\end{gather}
where $O^a(\omega)$ is constructed from the boundary fields~$\zeta(\vec x)$ at $x_0=0$.  To achieve a high sensitivity, the wave functions are constructed so as to approximate the ground and first excited state in the pseudoscalar channel.  For this purpose we examine the boundary-to-boundary correlator $f_1$ between a source term $O^a(\omega)$ at $x_0=0$ and another one, $O'^a(\omega')$, at $x_0=T$:
\begin{gather}
  f_1(\omega',\omega) = -\frac{1}{3L^6}\expectationvalue{O'^a(\omega')O^a(\omega)}.
\end{gather}
We compute its values for all combinations of the following three basis wave functions $\omega_1$, $\omega_2$, $\omega_3$:
\begin{gather}
  \bar\omega_1(r) = \ee^{-r/r_0}, \qquad \bar\omega_2(r) = r\cdot\ee^{-r/r_0}, \qquad \bar\omega_3 = \ee^{-r/(2r_0)}, \\
  \omega_i(\vec x) = N_i\sum_{\vec n\in\mathbb Z^3}\bar\omega_i(|\vec x-\vec nL|),
\end{gather}
where $N_i$ is a proper normalization factor and $r_0$ is a physical scale, which we choose to be $L/6$.

The first and second eigenvector $\eta^{(0)}$ and $\eta^{(1)}$ of the $3\times3$ matrix $[f_1(\omega_i',\omega_j)]_{i,j=1,2,3}$ are used to approximate the wave functions of the ground and first excited state:
\begin{align}
  \omega_{\pi^{(0)}} & \approx \sum_{i=1}^3\eta^{(0)}_i\omega_i, & \omega_{\pi^{(1)}} & \approx \sum_{i=1}^3\eta^{(1)}_i\omega_i.
\end{align}

With these choices, our definition of $\ca$ reads
\begin{gather}
  \label{e:ca_definition}
  \ca = -\frac{1}{a}\cdot\frac{r(x_0;\omega_{\pi^{(1)}})-r(x_0;\omega_{\pi^{(0)}})}{s(x_0;\omega_{\pi^{(1)}})-s(x_0;\omega_{\pi^{(0)}})}, \\
  r(x;\omega) = \frac{\frac{1}{2}(\partial_0+\partial^*_0)\fa(x_0;\omega)}{2\fp(x_0;\omega)}, \qquad
  s(x;\omega) = \frac{\partial_0\partial^*_0\fp(x_0;\omega)}{2\fp(x_0)},
\end{gather}
where the wave functions $\omega_{\pi^{(0)}}$ and $\omega_{\pi^{(1)}}$ determine the states $\beta$ and $\delta$ from eq.~\eqref{e:ca_definition_general}, whereas $\ket{\alpha}$ and $\ket{\gamma}$ have vacuum quantum numbers.  To complete our specification of the improvement condition, we still have to choose a value for $x_0$.

\section{Simulations}

\begin{table}[b]
  \centering
  \begin{tabular}{lllllll}
    \toprule
    $L/a$ & $T/a$ & $\beta$ & $\kappa$ & $am_\PCAC$ & $N_\mathrm{tr}$ & acc. rate \\
    \midrule
    12 & 19 & 3.3   & 0.13652 & $-0.0002(9)$ & 1000 & 0.96 \\
    16 & 23 & 3.512 & 0.13703 & $+0.0056(3)$ & 4096 & 0.93 \\
    20 & 29 & 3.676 & 0.1368  & $+0.0145(5)$ & \multicolumn{2}{l}{\emph{in progress}} \\
    24 & 35 & 3.810 & 0.13712 & $-0.00279(12)$ & 2604 & 0.93 \\
    \bottomrule
  \end{tabular}
  \caption{Simulation parameters, number of trajectories and the average acceptance.  $am_\PCAC$ is computed from the correlation functions projected to the approximate ground state, using the one-loop $\ca$ from \cite{Aoki:1998qd}.}
  \label{t:simulation_parameters}
\end{table}

As in the determination of $\csw$ in \cite{Bulava:2013cta}, the gauge configurations are created by the openQCD code (version~1.2 \cite{Luscher:2012av,website:openqcd}), which can also deal with SF boundary conditions.  It employs the HMC algorithm \cite{Duane:1987de} with frequency splitting of the quark determinant for a doublet out of the three dynamical quarks \cite{Hasenbusch:2001ne}.  We have added a small twisted mass regulator at $L/a=12$ \cite{Luscher:2008tw} but have not found it necessary for the stability of the other simulations.  The action of the third dynamical quark is simulated via the RHMC algorithm \cite{Clark:2006fx} with eight or nine poles.  The twisted mass at $L/a=12$ and the rational approximation are corrected afterwards with stochastically estimated reweighting factors.  The length of one trajectory is $\tau=2$.

Our simulations have $L/a=12$, 16, 20 and 24.  The time extent is such that~$T/L\approx3/2$.  We have picked this ratio, because we plan to reuse the configurations for a determination of the renormalization factor~$\za$.  Moreover, we set $\theta$ and the background field to zero.  A summary of the simulation parameters can be found in table~\ref{t:simulation_parameters}.

The parameters $\beta$ and $\kappa$ are adjusted so that they lie on a LCP.  $\beta$ is fixed by choosing the initial value $\beta=3.3$ for the coarsest lattice with $L/a=12$ and keeping its physical length~$L$ fixed in all simulations.  The $\beta$-values for the finer lattices are estimated by the perturbative formula
\begin{align}
  \frac{a(g_0^2)}{a(g_0'^2)} & = \ee^{-\left(g_0^{-2}-g_0'^{-2}\right)/(2b_0)}\left(g_0^2/g_0'^2\right)^{-b_1/(2b_0^2)}\times\left[1+q\left(g_0^2-g_0'^2\right)+\order{g_0'^4}\right], & g_0 & < g_0'.
\end{align}
Since the 3-loop contribution to $q$ for our action is not known, we include the universal parts only.

In order to completely define a LCP we must also fix the quark masses.  For the purpose of defining $\ca$ we can work with massless quarks, which is possible for SF boundary conditions.  In practice it is sufficient to employ three mass-degenerate quarks with $\kappa$-values such that the PCAC mass is reasonably small.  Based on experience from \cite{Bulava:2013cta}, where results on $\csw$ were found to be insensitive to violations of this condition within $|am_\PCAC|<0.015$, we take the same limit as a guideline for the tuning of our simulations.  $\kappa$ and the PCAC masses are also listed in table \ref{t:simulation_parameters}.

To check for the size of $\order{a}$ ambiguities, the generation of a further ensemble at $L/a=12$ and $T/a=17$ is under way, after tuning and tests were done with $T/a=19$, which facilitated parallelization.  Due to similar reasons, the $L/a=20$ simulations are still in progress.

\begin{figure}[tb]
  \centering
  \includegraphics[width=\textwidth]{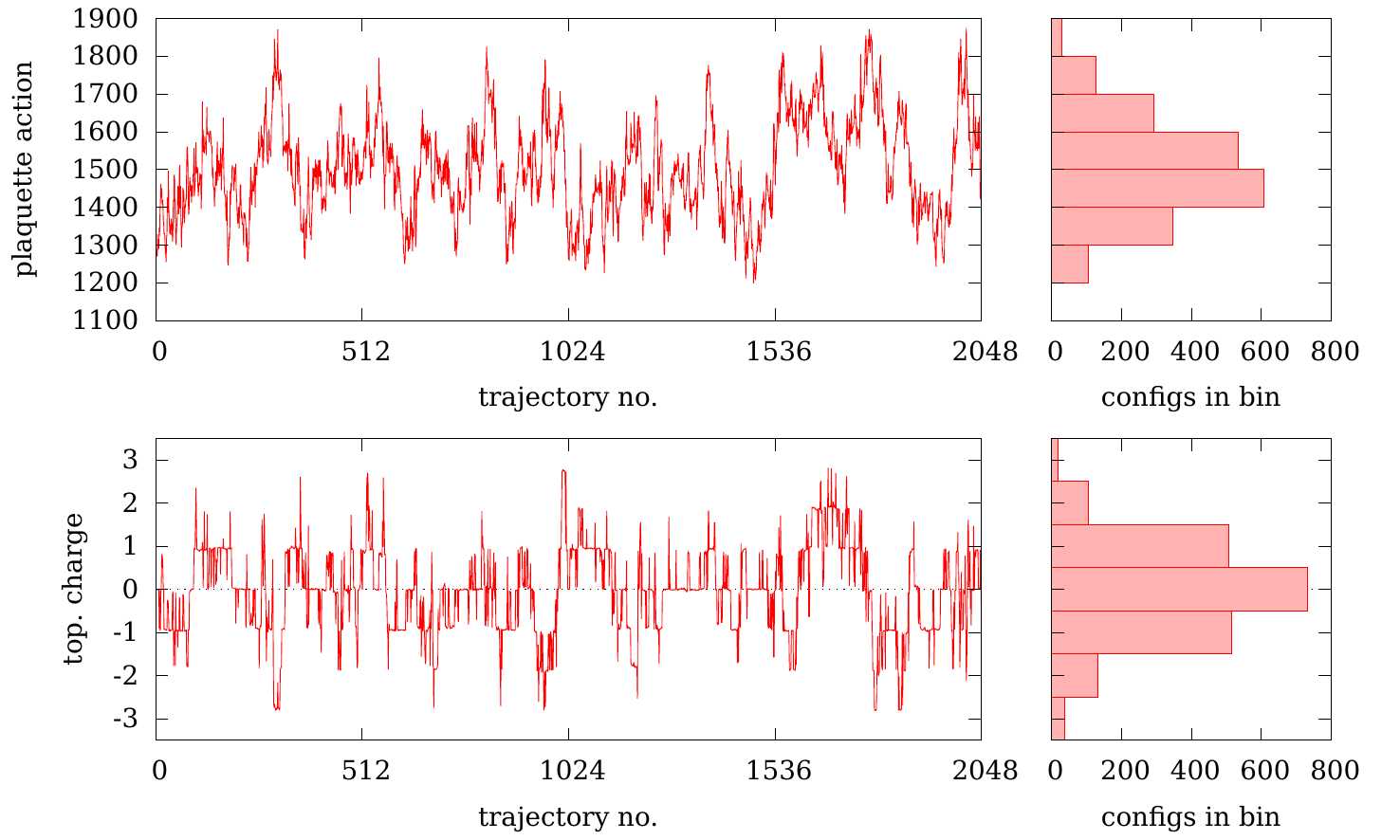}
  \caption{First 2048 trajectories of the histories and distributions of the smoothed plaquette action and the topological charge at Wilson flow time $t=(0.35\cdot L)^2/8$ of the simulation at $L/a=16$ and $\kappa=0.13703$.}
  \label{f:histories}
\end{figure}

As an example, figure~\ref{f:histories} shows an excerpt of the histories of the smoothed Wilson plaquette action and the topological charge of a simulation at $L/a=16$ and $\kappa=0.13703$.  They are taken at the Wilson flow time $t=(c\cdot L)^2/8$ with $c=0.35$ (cf.\ \cite{Fritzsch:2013je}).

\section{Preliminary Results}
\label{s:results}

\begin{figure}[tb]
  \centering
  \includegraphics[trim=0.6cm 8.3cm 1.3cm 9.5cm,width=0.49\textwidth]{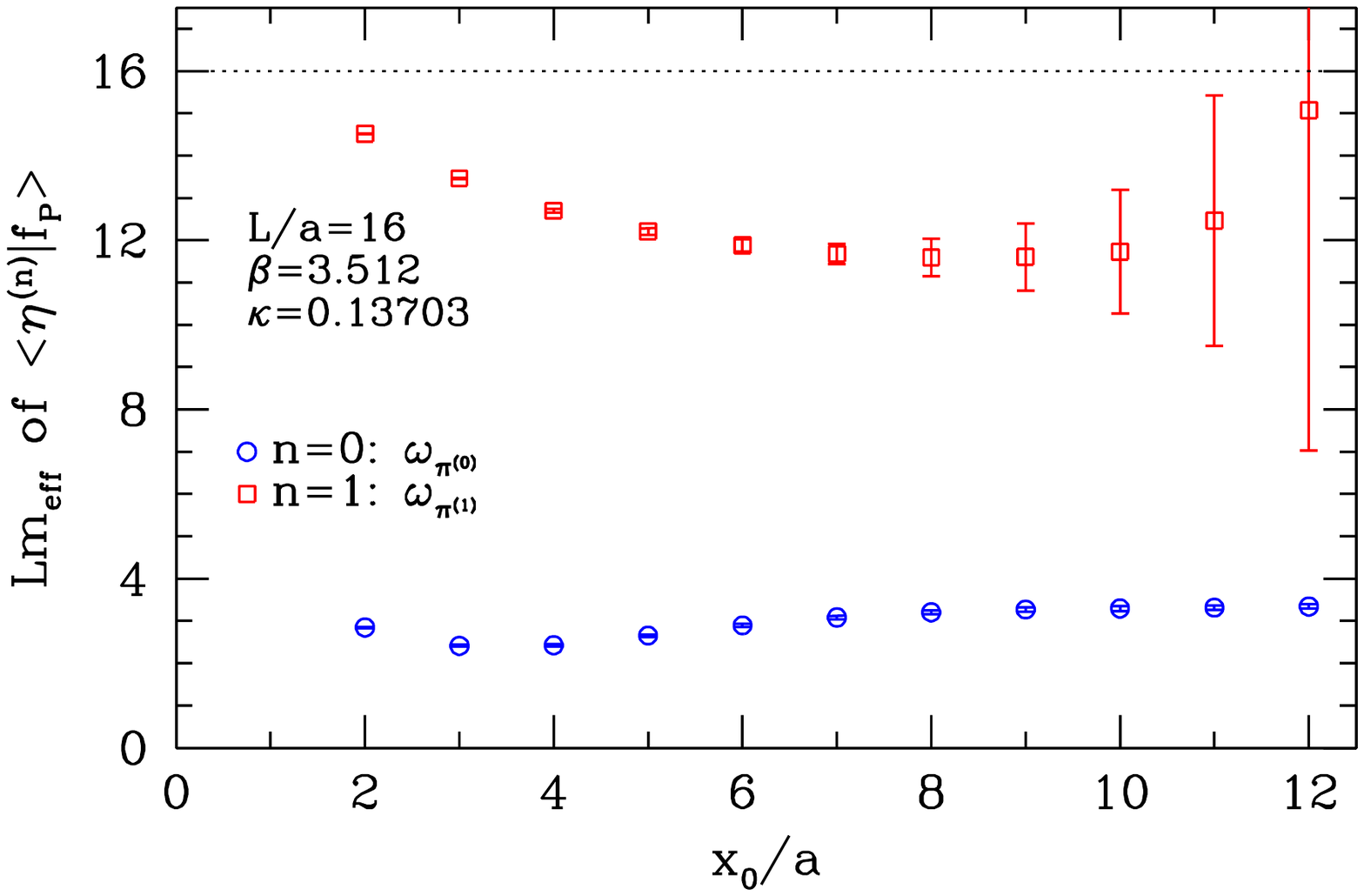}
  \hfill
  \includegraphics[trim=0.6cm 8.3cm 1.3cm 9.5cm,width=0.49\textwidth]{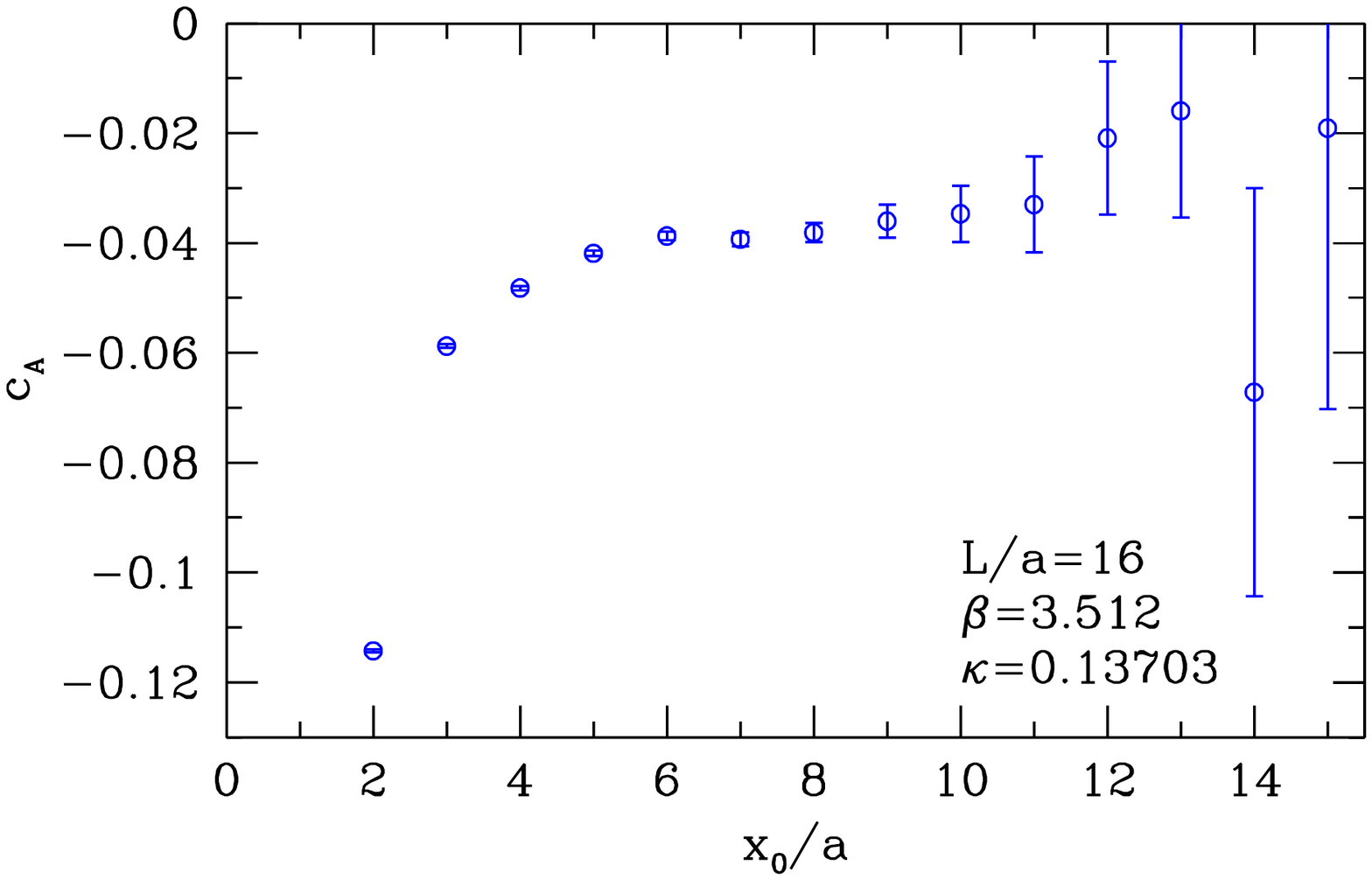}
  \caption{\emph{Left:} Effective masses computed from $\fp$ with wave functions $\omega_{\pi^{(0)}}$ and $\omega_{\pi^{(1)}}$ for $L/a=16$ and $\kappa=0.13703$.  \emph{Right:} $\ca(x_0)$ for $L/a=16$ and $\kappa=0.13703$.}
  \label{f:effective_masses}
  \label{f:ca_versus_x0}
\end{figure}

\begin{figure}[tb]
  \centering
  \newbox\cabox
  \setbox\cabox\hbox{\includegraphics[trim=0.7cm 8.3cm 1.1cm 9.5cm,width=0.5\textwidth]{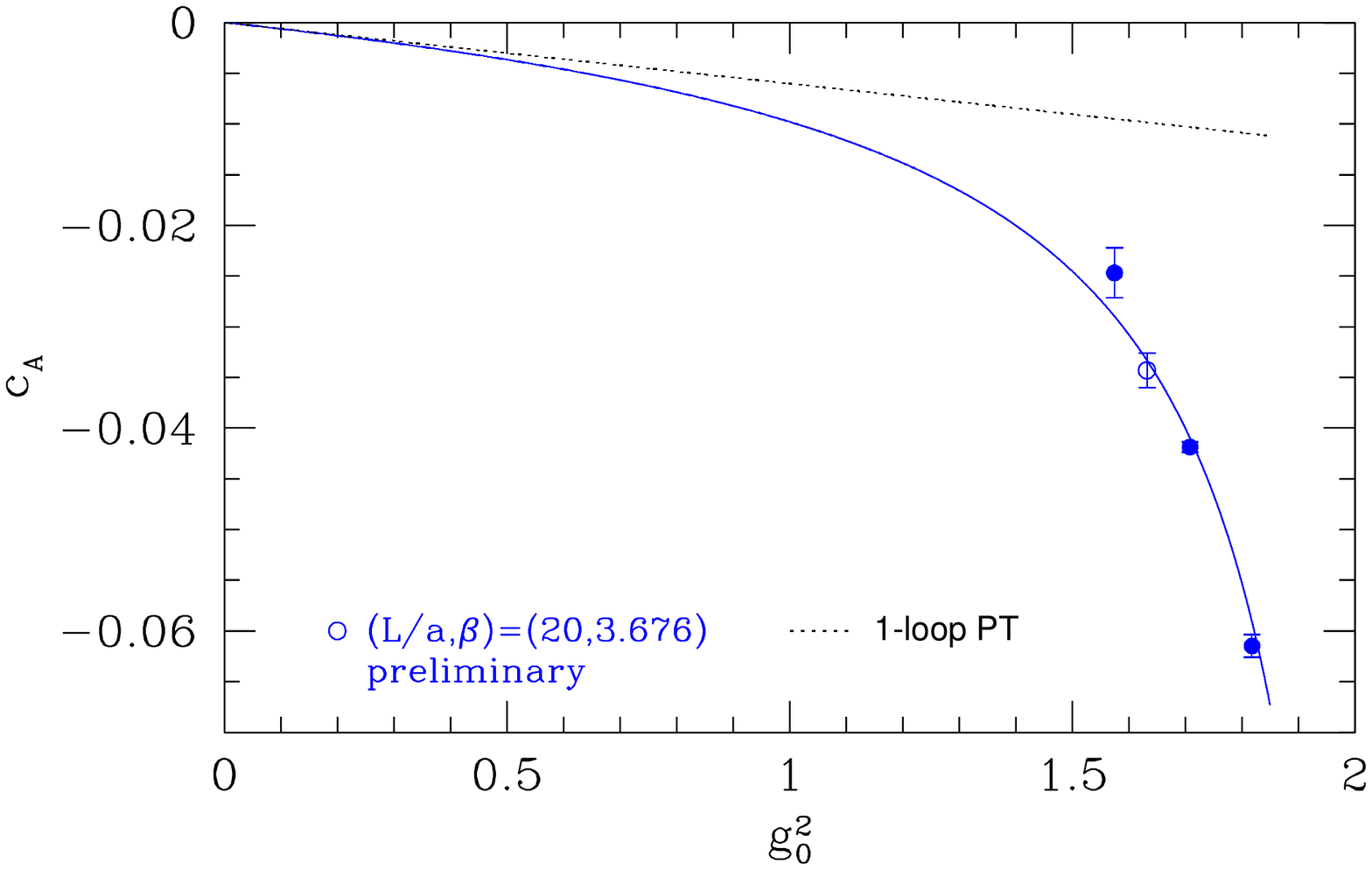}}
  \noindent\lower0.5\ht\cabox\box\cabox
  \hfill
  \begin{tabular}{lll}
    \toprule
    $L/a$ & $\beta$ & $\ca$ \\
    \midrule
    12 & 3.3   & $-0.0615(11)$ \\
    16 & 3.512 & $-0.0419(5)$ \\
    20 & 3.676 & $-0.0343(17)$ \\
    24 & 3.810 & $-0.025(2)$ \\
    \bottomrule
  \end{tabular}
  \hspace*{\stretch{1}}
  \caption{$\ca$ determined at $x_0=L/3$; the graph shows the data plotted against $g_0^2=6/\beta$ as well as the function $f_\ca$ fitted to them (the open data point from $L/a=20$ is preliminary and only included for illustration); the dashed line represents the one-loop asymptotics of $\ca$.}
  \label{f:ca_table}
  \label{f:ca_plot}
\end{figure}

The measurements of the SF correlation functions are performed on every fourth configuration only, in order to reduce autocorrelations.  So far, errors are estimated by a binned Jackknife analysis.

The first step of the analysis is to determine the eigenvectors $\eta^{(0)}$ and $\eta^{(1)}$ of the matrix $[f_1(\omega_i',\omega_j)]_{i,j=1,2,3}$.  The normalized eigenvectors have a well-defined continuum limit along a LCP, as long as the wave functions depend on physical scales only.  In agreement with this expectation and the findings in \cite{DellaMorte:2005se,Kaneko:2007wh}, we observe no strong dependence on the lattice spacing for them.  Therefore, we fix the vectors for once at their values from the analysis of $L/a=16$ and $\kappa=0.13703$, which can be regarded as part of the improvement condition.  These eigenvectors are $\eta^{(0)}=(0.5317(3),0.5977(1),0.6000(2))$ and $\eta^{(1)}=(0.843(5),-0.31(6),-0.44(6))$.

With these vectors we project the correlation functions to the approximate ground and first excited state, $f_{\mathrm A/\mathrm P}(x_0;\omega_{\pi^{(0)}})$ and $f_{\mathrm A/\mathrm P}(x_0;\omega_{\pi^{(1)}})$.  $\fp$ is used to compute the effective masses of both states.  Representative results for $L/a=16$ and $\kappa=0.13703$ are shown in figure~\ref{f:effective_masses}.  The two states are clearly seen to be separated up to $x_0\approx12a$.  A plot of the local $\ca$ obtained according to \eqref{e:ca_definition} for this data set can be found in figure~\ref{f:ca_versus_x0} on the right.  For the final definition of $\ca$ we choose $x_0=L/3$, as it seems to be already in the asymptotic regime but still has a good signal-to-noise ratio.  The resulting $\ca$-values from our present analysis are compiled in the table in figure~\ref{f:ca_table}.

In figure~\ref{f:ca_plot} the results for $\ca$ are plotted against $g_0^2=6/\beta$.  The solid line is an interpolation of the data based on the functional form
\begin{align}
  f_\ca(g_0^2) & = -0.006033\cdot g_0^2\cdot\frac{1+p_1\cdot g_0^2}{1+p_2\cdot g_0^2},
\end{align}
which is constrained to the one-loop value from \cite{Aoki:1998qd}.  The resultant parameters are $p_1=-0.15(2)$ and $p_2=-0.476(6)$.  The $\ca$ from $L/a=20$ is not included, since simulations are still in progress.

\section{Outlook}

For the final analysis, we will not only include the still ongoing simulations at $L/a=20$ and $L/a=12$ with $T/a=17$, but also increased statistics for the other $\beta$-values.  Moreover, we also want to investigate the influence of small deviations from the constant-physics condition on our results.

\section{Acknowledgments}

We want to thank Rainer Sommer and Stefan Schaefer for helpful discussions.  This work is supported by the grant HE~4517/3-1 (J.~H. and C.~W.) of the Deutsche Forschungsgemeinschaft.  Furthermore, we gratefully acknowledge the computer resources provided by DESY, Zeuthen (PAX Cluster), CERN and the ZIV of the University of Münster (PALMA HPC cluster).

\bibliographystyle{bibstyles/JHEP_new_notitles}
\bibliography{proceedings}

\end{document}